\documentstyle[pra,preprint,aps]{revtex}
\begin{document}
\draft
\title{Physical limitations on quantum nonlocality in the detection of gamma photons
emitted from positron/electron annihilation}
\author{Victor D. Irby}
\address{Department of Physics, University of South Alabama, Mobile, Al 36688-0002}
\date{\today}
\maketitle
\begin{abstract}
Recent experimental measurements of the time interval between
detection of the two photons emitted in positron/electron
annihilation have indicated that collapse of the spatial part of
the photon's wavefunction, due to detection of the other photon,
\textit{does not occur}. Although quantum nonlocality actually
occurs in photons produced through parametric down-conversion, the
recent experiments give strong evidence \textit{against}
measurement-induced instantaneous spatial-localization of
high-energy gamma photons. A new quantum-mechanical analysis of
the EPR problem is presented which may help to explain the
observed differences between photons produced through parametric
down-conversion and photons produced through positron/electron
annihilation. The results are found to concur with the recent
experiments involving gamma photons.
\end{abstract}


Quantum non-locality in measurements involving polarization and
two-photon interference of correlated photons has been
experimentally confirmed at many independent laboratories [1-4].
It has been generally postulated that non-local effects may also
occur in regard to the spatial wavefunctions of the emitted
photons.  As an example, detection of one of the photons produced
in parametric down conversion is predicted to cause
``instantaneous" localization of the other photon, subsequently
eliminating any uncertainty in the time of arrival of the second
photon at a second detector. Experimental support of this
prediction has been reported by Hong \textit{et al.} \cite{Hong}.
The two-photon interference method utilized in Ref. \cite{Hong}
indicates that the minimum time uncertainty, in the time interval
between detection of the two down-converted photons, is less than
$100 fs$. This uncertainty in time is \textit{much less} than the
coherence time of the initial pump photons, which subsequently
gives strong indication of nonlocal collapse of the photon
wave-function.

One may expect to observe similar nonlocal effects involving
photons emitted from positron/electron annihilation.  Recent
high-resolution measurements of the time interval between the two
photons emitted in positron/electron annihilation have been
carried out by Irby \cite{Irby}.  The results of the measurements
indicate that the \textit{absolute minimum} uncertainty in
detection time between arrival of the two photons is $\Delta
t_{QM}=117\pm 9 \,ps$, which surprisingly, agrees with the
lifetime of positrons in bulk sodium ($119\,ps$) predicted by
quantum electrodynamics \cite{Sakurai} \cite{Charlton}. Although
nonlocal effects are observed to occur in the case of down
converted photons, the experimental results give strong evidence
\textit{against} the instantaneous spatial-localization of gamma
photons emitted from annihilation events.

In this paper, we present a quantum-mechanical analysis of the
time interval between detection of correlated photons.  The
analysis is basically the same as that first presented by
Einstein, Podolsky, and Rosen (EPR) in 1935 \cite{Einstein}.  The
main difference, however, is that we include time dependence in
the quantum wavefunctions and also take into account restrictions
on photon momenta due to energy conservation.

As in the original EPR paper, we assume that the total momentum
before the particles interact (or are emitted) is zero. In
addition, we assume that the particles interact at times $t<0$.
The total wavefunction can then be written (for $t\geq 0$)

\begin{equation}
\Psi(x_{1},x_{2},t)\,\,=\,\,\int_{-\infty}^{\infty}\psi_{p}(x_{2},
t)\,u_{p}(x_{1}, t)\,dp\, ,\label{eq1}
\end{equation}

\noindent where $u_{p}(x_{1}, t)$ are eigenstates of particle
one's momentum and energy

\begin{equation}
u_{p}(x_{1},t)\,\,=\,\,e^{ipx_{1}/\hbar}\,e^{-iEt/\hbar}\,\,.\label{eq2}
\end{equation}

\noindent In order to conserve momentum, let us also assume

\begin{equation}
\psi_{p}(x_{2},t)\,\,=\,\,e^{-ip(x_{2}\,+\,x_{o})/\hbar}\,e^{-iEt/\hbar}\,\,,\label{eq3}
\end{equation}

\noindent which are eigenstates of particle two's momentum and
energy. ($x_{o}$ is an arbitrary constant introduced in the
original EPR paper.  In this case, however, since we are including
the explicit time dependence, we will set $x_{o}=0$). Note that if
a measurement of particle one's momentum yields a value of $p$,
the total wavefunction collapses to

\begin{equation}
\Psi (x_{1},x_{2},t)\,\,=\,\,\psi_{p}(x_{2}, t)\,u_{p}(x_{1},
t)\,\,,\label{eq4}
\end{equation}

\noindent which has momentum eigenvalues of $p$ and $-p$
respectively for particles one and two.  Before any measurement
takes place, the total wavefunction is thus given by

\begin{equation}
\Psi (x_{1},x_{2},t)\,\,=\,\,\int_{-\infty}^{\infty}\,
e^{ip(x_{1}\,-\,x_{2})/\hbar}\,e^{-i2Et/\hbar}\,dp\,\,.\label{eq5}
\end{equation}

\noindent The total wavefunction can also be written in terms of
\textit{instantaneous} position eigenstates $v_{x}(x_{1},t)$ of
particle one

\begin{equation}
\Psi(x_{1},x_{2},t)\,\,=\,\,\int_{-\infty}^{\infty}\phi_{x}(x_{2},
t)\,v_{x}(x_{1},t)\,dx\, ,\label{eq6}
\end{equation}

\noindent where particle two's wavefunction $\phi_{x}(x_{2}, t)$
has yet to be specified.  Since the position eigenstates of
particle one, measured at time $t$, are Dirac delta-functions
$v_{x}(x_{1}, t)=\delta (x-x_{1})$, with eigenvalues $x_{1}$, Eq.
6 reduces to

\begin{equation}
\Psi(x_{1},x_{2},t)\,\,=\,\,\phi_{x_{1}}(x_{2}, t)\, ,\label{eq7}
\end{equation}

\noindent where $x_{1}$ now represents an eigenvalue measured at
time $t$. Therefore, the spatial wavefunction of particle two is
dependent on the position measurement $x_{1}$ of particle one:

\begin{equation}
\phi_{x_{1}}(x_{2},t)\,\,=\,\,\int_{-\infty}^{\infty}\,
e^{ip(x_{1}\,-\,x_{2})/\hbar}\,e^{-i2Et/\hbar}\,dp\,\,.\label{eq8}
\end{equation}

Let us first consider the case where the particles have non-zero
rest mass.  If particle one's position $x_{1}$ is measured at
$t=0$ , Eq. 8 reduces to

\begin{eqnarray}
\phi_{x_{1}}(x_{2},0)\,\,=\,\,\int_{-\infty}^{\infty}\,
e^{ip(x_{1}\,-\,x_{2})/\hbar}\,dp \nonumber, \\
\phi_{x_{1}}(x_{2},0)\,\,=\,\,\hbar \,\delta
(x_{1}\,-\,x_{2})\,\,,\label{eq9}
\end{eqnarray}

\noindent resulting in particle two being localized at
$x_{2}=x_{1}$ (which is the same result as presented in the
original EPR paper with $x_{o}=0$). If particle one's position
$x_{1}$ is measured at a time other than zero, particle two's
wavefunction at the measurement time $t$ is then explicitly given
by

\begin{equation}
\phi_{x_{1}}(x_{2},t)\,\,=\,\,\int_{-\infty}^{\infty}\,
e^{ip(x_{1}\,-\,x_{2})/\hbar}\,e^{-i2\sqrt{(pc)^{2} +
(mc^{2})^2}t/\hbar}\,dp\,\,.\label{eq10}
\end{equation}

\noindent  Thus, if particle one's position $x_{1}$ is measured at
a time other than $t=0$, particle two \textit{will not be
localized}. Particle two is only localized at one instant, namely
$t=0$. Furthermore, regardless of \textit{when} particle two is
localized, particle two's wavefunction will always immediately and
rapidly disperse as time progresses.

In contrast, since $E=pc$ for photons, dispersion no longer
exists. The spatial wavefunction for photons is given by

\begin{eqnarray}
\phi_{x_{1}}(x_{2},t)\,\,=\,\,\int_{-\infty}^{\infty}\,
e^{ip(x_{1}\,-\,x_{2}\,-\,2ct)/\hbar}\,dp \nonumber,
\\
\phi_{x_{1}}(x_{2},t)\,\,=\,\,\hbar\,\delta
(x_{1}\,-\,x_{2}\,-\,2ct).\label{eq11}
\end{eqnarray}

\noindent  Thus, after measurement of photon one's location
$x_{1}$ at time $t$, photon two is instantaneously localized at
$x_{2}=x_{1}-2ct$.  In contrast with particles of non-zero rest
mass, once photon two is localized, it will remain localized
(propagating at c).

The result, given in Eq. 11 above, contradicts the experimental
measurements. Localization of the second photon should
\textit{eliminate} any uncertainty in arrival time between the two
photons. This glaring contradiction between theory and experiment
can, however, be elieviated by properly taking into account
necessary restrictions on photon momenta.

For the case of positron/electron annihilation, the emitted
photons are restricted to a small range of possible momenta
$\Delta p$ centered at $p=mc$ in order for energy to be conserved.
In order to take into account conservation of energy, let
$\,|f(p)|^{2}dp\,$ be the probability of photons having momenta
between $p$ and $p+dp$. The total wavefunction is then given by

\begin{equation}
\Psi(x_{1},x_{2},t)\,\,=\,\,\int_{-\infty}^{\infty}f(p)\,\,\psi_{p}(x_{2},
t)\,u_{p}(x_{1}, t)\,\,dp\,\,.\label{eq12}
\end{equation}

\noindent  Once photon one is detected, photon two's wavefunction
is then given by

\begin{equation}
\phi_{x_{1}}(x_{2},t)\,\,=\,\,\int_{-\infty}^{\infty}\,f(p)\,
e^{ip(x_{1}\,-\,x_{2}\,-\,2ct)/\hbar}\,dp\,\,,\label{eq13}
\end{equation}

\noindent which is no longer equal to a Dirac delta-function. As
Eq. 13 indicates, restrictions on emitted photon momenta
\textit{prohibits} instantaneous and complete localization of the
second photon.

The prohibition on nonlocality indicated above may also be
described in terms of partial entanglement \cite{Barnett}
\cite{White}. As is well known in the quantum-optics community, if
a particular observable is subject to physical restrictions, any
other conjugate observable, associated with a non-commuting
operator, will exhibit a corresponding restriction in terms of
nonlocality. This can be more easily shown in terms of spin
measurements.

Let us assume that two particles are emitted such that the total
spin wave function, measured along the $x$ axis, is given by

\begin{equation}
\Psi_{x}\,\,=\,\,a|\uparrow\rangle_{1}|\downarrow\rangle_{2}\,\,-\,\,
b|\downarrow\rangle_{1}|\uparrow\rangle_{2}\,\,,\label{eq14}
\end{equation}

\noindent where $a^{2} + b^{2}=1$.  If $a=b$, then individual spin
measurements (along $x$) for either particle are unrestricted and
completely uncertain.  If $a\neq b$, there exists partial
restriction.  If either $a$ or $b$ is equal to zero, spin
measurements are completely restricted.  Note that the above
wavefunction will always exhibit maximum nonlocality (for
measurements along $x$) regardless of the values of $a$ and $b$. A
measurement of $|\uparrow\rangle_{1}$ will always yield
$|\downarrow\rangle_{2}$.

Particle spin along the $z$ axis, however, is a conjugate
observable.  For spin measurements along the $z$ axis, the total
wavefunction can easily be shown to be

\begin{equation}
\Psi_{z}\,\,=\,\,\frac{1}{\sqrt{2}}\,|\uparrow\rangle_{1}\,\,\bigg{\{}\frac{(a-b)}{\sqrt{2}}|\uparrow\rangle_{2}\,-\,
\frac{(a+b)}{\sqrt{2}}|\downarrow\rangle_{2}\bigg{\}}\,\,\,\,+\,\,\,\,\frac{1}{\sqrt{2}}\,|\downarrow\rangle_{1}\,\,\bigg{\{}\frac{(a+b)}{\sqrt{2}}|\uparrow\rangle_{2}\,-\,
\frac{(a-b)}{\sqrt{2}}|\downarrow\rangle_{2}\bigg{\}}\,.\label{eq15}
\end{equation}

If $a=b$, observables associated with spin measurements along $x$
are unrestricted.  For spin measurements along $z$, Eq. 15 reduces
to (with $a=b$)

\begin{equation}
\Psi_{z}\,\,=\,\,-\frac{1}{\sqrt{2}}|\uparrow\rangle_{1}|\downarrow\rangle_{2}\,\,+\,\,
\frac{1}{\sqrt{2}}|\downarrow\rangle_{1}|\uparrow\rangle_{2}\,\,.\label{eq16}
\end{equation}

\noindent In this case, spin measurements along the $z$ axis
exhibit maximum nonlocality, or maximum entanglement.

In contrast, let us assume $a=0.8$ and $b=0.6\,$  In this case,
spin measurements along $x$ are only slightly restricted.  For
spin measurements along the z axis, Eq. 15 reduces to

\begin{equation}
\Psi_{z}\,\,=\,\,\frac{1}{\sqrt{2}}\,|\uparrow\rangle_{1}\,\,\bigg{\{}0.141|\uparrow\rangle_{2}\,-\,
0.989|\downarrow\rangle_{2}\bigg{\}}\,\,\,\,+\,\,\,\,\frac{1}{\sqrt{2}}\,|\downarrow\rangle_{1}\,\,\bigg{\{}0.989|\uparrow\rangle_{2}\,-\,
0.141|\downarrow\rangle_{2}\bigg{\}}\,.\label{eq17}
\end{equation}

\noindent In this case, spin measurements along $z$ no longer
exhibit maximum nonlocality.  If a measurement of particle one's
spin yields $|\uparrow\rangle_{1}$, only $98\%$ of the time will
particle two yield $|\downarrow\rangle_{2}$.

On the other extreme, if $b=0$, the observables associated with
spin measurements along $x$ are maximally restricted.  For spin
measurements along the $z$ axis, Eq. 15 reduces to (with $b=0$)

\begin{equation}
\Psi_{z}\,\,=\,\,\frac{1}{\sqrt{2}}\,|\uparrow\rangle_{1}\,\,\bigg{\{}\frac{1}{\sqrt{2}}|\uparrow\rangle_{2}\,-\,
\frac{1}{\sqrt{2}}|\downarrow\rangle_{2}\bigg{\}}\,\,\,\,+\,\,\,\,\frac{1}{\sqrt{2}}\,|\downarrow\rangle_{1}\,\,\bigg{\{}\frac{1}{\sqrt{2}}|\uparrow\rangle_{2}\,-\,
\frac{1}{\sqrt{2}}|\downarrow\rangle_{2}\bigg{\}}\,.\label{eq18}
\end{equation}

\noindent In this case, measurement of particle one's spin
\textit{does not in any way influence} the measurement of the spin
of particle two.  Nonlocality is erased.

As the above analysis indicates, measurements involving a
particular observable \textit{may or may not} exhibit nonlocality,
depending upon the degree of physical restraints that may exist on
conjugate observables.  In the case of photons being emitted from
positron/electron annihilation, photon momenta are, for all
practical purposes, \textit{maximally restricted}. This then
essentially eliminates nonlocality in the conjugate position
observables.  Therefore $\triangle t_{QM} \neq 0$.  (However,
partial entanglement still, nonetheless, exists. In the limit of
well defined momenta, $\triangle t_{QM} \rightarrow \infty$ for
the case of no entanglement).  In striking contrast, parametric
down-converted photons exhibit a \textit{much stronger
correlation} in time than that of gamma photons. This may be
attributed to the fact that down-converted photons possess much
larger uncertainties in emission energy than those of gamma
photons.

%
%


\begin{references}

\bibitem{Aspect}A.~Aspect, J.~Dalibard, and G.~Roger,
Phys.~Rev.~Lett., {\bf49}, 1804 (1982).

\bibitem{Franson1}J.~D.~Franson, Phys.~Rev.~A, {\bf44}, 4552
(1991).

\bibitem{Franson2}J.~D.~Franson, Phys.~Rev.~Lett.~{\bf62}, 2205
(1989).

\bibitem{Hong}C.~K.~Hong, Z.~Y.~Ou, and L.~Mandel,
Phys.~Rev.~Lett.~{\bf59}, 2044 (1987).



\bibitem{Irby}V.~D.~Irby, (submitted for publication, Meas. Sci. and
Tech.)
[http://www.southalabama.edu/physics/faculty/irby.html].

\bibitem{Sakurai}J.~J.~Sakurai, in
$\textit{Advanced Quantum Mechanics}$, (Addison-Wesley, Reading
Mass. 1977), pg. 216.

\bibitem{Charlton}M.~Charlton, J.~W.~Humberston, in
$\textit{Positron Physics}$, (Cambridge University Press, 2001),
pg.264.

\bibitem{Einstein}A.~Einstein, B.~Podolsky, and N.~Rosen,
Phys.~Rev.~{\bf47}, 777 (1935).

\bibitem{Barnett}S.~M.~Barnett and S.~J.~D.~Phoenix, Phys.~Rev.~A,
{\bf40}, 2404 (1989).

\bibitem{White}A.~G.~White, D.~F.~V.~James, W.~J.~Munro, and
P.~G.~Kwiat, Phys.~Rev.~A, {\bf65}, 012301 (2001).


\end{references}
\end{document}